\documentclass[12pt]{iopart}
\usepackage{amssymb,latexsym,mathrsfs}
\usepackage{amsfonts}
\usepackage{graphicx}
\usepackage{color}    
\usepackage{textcomp}

\newcommand{\beq}{\begin{equation}}
\newcommand{\eeq}{\end{equation}}
\newcommand{\bearr}{\begin{array}}
\newcommand{\enarr}{\end{array}}

\begin{document}

\title{Multibalance conditions in nonequilibrium steady states}
\author{Indranil Mukherjee}
\address{Condensed Matter Physics Division, Saha Institute of Nuclear Physics, HBNI,
1/AF Bidhan Nagar, Kolkata, 700064 India.}
\ead{indranil.mukherjee@saha.ac.in}

\begin{abstract}
We study a new balance condition multibalance to obtain the nonequilibrium steady states of a class of nonequilibrium lattice models on a ring where a particle hops from a particular site to its nearest and next nearest  neighbours. For the well-known zero range process (ZRP)  with asymmetric hop rates, with this balance condition, we obtain the  conditions on hop rates that lead to a factorized steady state (FSS). We  show that this balance condition  gives the cluster-factorized steady state (CFSS) for finite range process (FRP) and other models. We also discuss the application  of multibalance condition to two  species FRP model with hop rates ranging up to $K$ nearest neighbours.
\end{abstract}
\maketitle

\section{Introduction}

Nonequilibrium steady states (NESS) \cite{DDS,book} differ from their equilibrium counterparts
which obey detailed balance  (DB) \cite{book_db_1,book_db_2}. DB ensures that there is no net flow of probability current among any pair of configurations leading to the well known
Gibbs-Boltzmann measure in its steady state. Such a generic measure is absent in nonequilibrium  situation raising a general question: \textit{how to obtain the steady states of non-equilibrium systems}. Obtaining non-equilibrium steady state measure has always been a subject of interest. In general, finding an exact NESS for any nonequilibrium dynamics is usually difficult.
It has been realized that exact solutions of steady state measures for certain non-
equilibrium systems and analytical calculation of observables bring much insight to
the understanding of the corresponding systems. In context of the exactly solvable
interacting non-equilibrium systems, there exist  a few successful models. The zero range process (ZRP) \cite{Spitzer, Evans_Braz, Evans_Hanney}, a lattice gas model without any hardcore exclusions, is perhaps the simplest of
them, which exhibits nontrivial static and dynamic properties in the steady state. It has found applications in different branches of science such as in describing phase separation criterion in driven lattice gases \cite{Kafri_Mukamel}, network
re-wiring \cite{Angel_Evans, PKM_JALAN}, statics and dynamics of extended objects \cite{Gupta_Berma_PKM, Daga_PKM}. etc. The corresponding steady states of the well studied model ZRP, can be achieved using pairwise balance  condition condition (PWB) \cite{PWB}, where one uses the following principle: for every transition  $C\to C'$  there exists a unique configuration $C''$ such that the flux coming to $C$ from $C'$ is
exactly balanced by the flux going from $C$ to $C ''$. A special case of pairwise balance condition is DB when $C''$ = $C'$. For PWB to hold, a necessary
condition is that the number of distinct incoming fluxes to any configuration must be
equal to the number of distinct outgoing fluxes from that configuration. A prototypical example of non-equilibrium processes is the totally asymmetric simple exclusion process (TASEP) on a ring. The steady state of the TASEP with open boundaries was obtained exactly by Derrida et. al in Ref. \cite{derrida__tasep_mpa} using matrix product ansatz (MPA),  where steady state weight of any configuration is
represented by a product of matrices  containing two non-commuting matrices, one for the occupied
site and the other for the vacant site.  After successful implementation of MPA in TASEP with open boundaries \cite{derrida__tasep_mpa, evans_mpa}, it has been used extensively to solve the
steady states of different generalizations of TASEP, e.g., TASEP with multiple species
of particles \cite{EVANS_MR}, TASEP with internal degrees of freedom \cite{UBASU_PKM_TASEP}; non-conserved systems with deposition, evaporation, coagulation-decoagulation like dynamics \cite{Hinrichsen}.  Another class of nonequilibrium model that has been studied recently is finite range process (FRP) \cite{FRP}, having cluster-factorized steady state (CFSS). The steady state of this model can  be achieved by both pairwise balance and h - balance condition \cite{FRP, asymm_FRP}  and there exists a finite dimensional transfer-matrix representation of the steady state. 

In this article we have tried to find other possible  balance conditions, beyond DB and PWB, to achieve NESS and refer to  this as multibalance (MB): for  every configuration $C,$  sum of the outgoing fluxes to one or more  configurations are  balanced by the sum of multiple  incoming fluxes. We have applied this balance conditions to few nonequilibrium models and obtained the exact steady states. We have studied the ZRP with directional asymmetry in two and three dimensions where we can get a factorized steady state (FSS) \cite{FSS_Evans_Zia, misanthrope_fss} 
with certain conditions on the hop rates using MB. We have studied the steady state condition of the FRP model with nearest neighbours and next nearest neighbours hopping for asymmetric rates and obtained the CFSS if the hop rates satisfy some specific conditions. We have extended this FRP model with two species of particles. This class of nonequilibrium lattice models can also  have  a cluster factorized steady state (CFSS) \cite{FRP}. For this model with directional asymmetry,  we achieved the condition  on hop rates that leads to a   pair-factorized steady state (PFSS). To this end,  we have also considered an interesting triangular lattice model, with particle hopping to its all nearest neighbours. Using MB, we have shown that one can have a pair-factorized steady states (PFSS) \cite{pfss} with certain conditions on the hop rates and discussed the way to find the observable for this system.

\section{Zero Range Process (ZRP) with directional asymmetry}

The zero range process (ZRP) is a model in which  many indistinguishable particles occupy sites on a lattice. Each lattice site may contain an integer number of particles and these particles hop between neighbouring sites with a rate that depends on the number of
particles at the site of departure.  The steady state of ZRP model in one dimension can be solved exactly for periodic and open boundary cases \cite{Evans_Hanney, open_ZRP_Levine, mpa_zrp}. ZRP in one dimension with asymmetric rates has been discussed recently in \cite{asymm_FRP}.
\subsection{ZRP in two dimensions with directional asymmetry}
\begin{figure}[h]
\begin{center}
\vspace*{.5 cm}
\includegraphics[scale = 0.25]{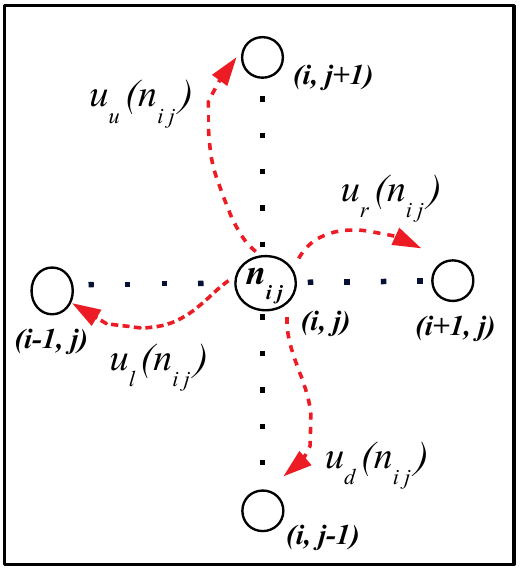}
\end{center}
\caption{ ZRP in two dimensions, a particle from site $(i,j)$ can hop to its right, left, up and down nearest neighbours with rates $u_{r}(n_{i,j})$, $u_{l}(n_{i,j})$, $u_{u}(n_{i,j})$ and $u_{d}(n_{i,j})$ respectively. $n_{i,j}$ is the number of particles at site $(i,j)$.}
\label{fig:2Dzrp}
\end{figure}

We consider a periodic ZRP lattice in two dimensions of size ($L \times L$). Each site $(i,j)$  with, $i = 1, 2, \cdots L, j = 1, 2, \cdots L$, can be vacant or it can be occupied by one or more particle $n_{i,j} \leq N$ $\left(N = \sum_{i=1}^{L} \sum_{j=1}^{L} n_{i,j} \right)$. A particle from any randomly chosen  site can hop to its nearest neighbours (right, left, up and down) with rates $u_{r}(n_{i,j})$, $u_{l}(n_{i,j})$, $u_{u}(n_{i,j})$ and $u_{d}(n_{i,j})$ respectively as shown in Fig. \ref{fig:2Dzrp}.

ZRP does not have an exact steady state when hop rates in all four directions are different. The model is solvable using (a) DB, when the rates $u_{r}(n) = u_{l}(n) = \alpha u(n)$ and $u_{u}(n) = u_{d}(n) = \beta u(n)$, where both $\alpha$ and $\beta$ are constants or  $u_{r}(n) = u_{l}(n) = u_{u}(n) = u_{d}(n) = u(n)$ and  (b) PWB, when all rates $u_{r}(n)$, $u_{l}(n)$,  $u_{u}(n)$ and $u_{d}(n)$ differ by a multiplicative constant {\it i.e.,} the ratios of the rates are independent of $n$,  assuming that the model evolves to a FSS 
\begin{equation} \label{eq:FSS_P}
 P(\{n_{i,j}\}) = \frac{1}{Q_{L,N}} \prod_{i=1,j=1}^{L} f(n_{i,j})~ \delta \left (\sum_{i=1,j = 1}^{L} n_{i,j} -N \right )
 \end{equation}
 where, the canonical partition function 
 \begin{equation}\label{eq:FSS_Q}
  Q_{L,N} = \sum_{\{n_{i,j} \}} \prod_{i=1,j=1}^{L} f(n_{i,j})~ \delta \left(\sum_{i=1, j = 1}^{L} n_{i,j} -N \right).
 \end{equation}
$N$ is the total number of particles and the density of the system $\rho = \frac{N}{L^{2}}$ is conserved by the dynamics. The steady state weight is defined as
\begin{equation} \label{eq:FSS_weight_steadty}
 f(n) = \prod_{\nu=1}^{n} u(\nu) ^{-1}.
\end{equation}
Is it possible to obtain an exact steady state using other flux cancellation schemes for this asymmetric hopping when the hop rates in all four directions are different, that increases the regime of solvability. We consider a new balance condition namely multibalance (MB).

\subsubsection{Multibalance (MB)}
\begin{figure}[h]
\vspace*{.5 cm}
\centering
\includegraphics[scale=0.3]{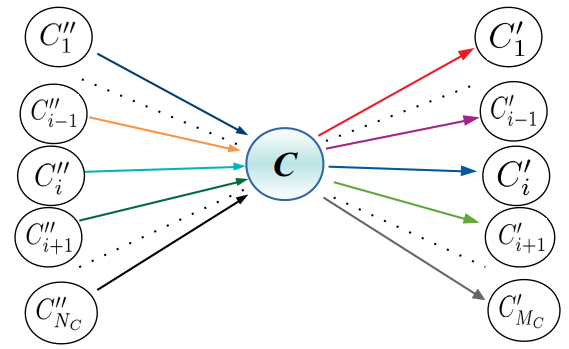} 
\caption{Multibalance (MB): fluxes are represented by arrows. Incoming fluxes to the configuration $C$ from a set of configurations  $\{ C''_{1}, \cdots,  C''_{N_{C}}\}$ are balanced with the outward fluxes from $C$ to the set of configurations $\{C'_{1}, ...,  C'_{M_{C}}\}$. Clearly, for $N_{C} = M_{C} = 1$, if   
$C''_{1} \neq C_{1}'$, MB reduces to PWB and  when $C''_{1} = C_{1}'$, it becomes DB  corresponds to the  equilibrium}
 \label{fig:multibalance}
\end{figure}

We define a generalized balance condition  in nonequilibrium systems such that a bunch of fluxes coming to the configuration $C$ from a  set of configurations  $\{ C''_{1}, \cdots, C''_{N_{C}}\}$ are balanced by the sum of out-fluxes from $C$  to a  set of configurations $\{C'_{1}, \cdots, C'_{M_{C}}\}$ in the configuration space.  Here $N_{C}$ is the total number of incoming fluxes for the set of configurations $\{ C''_{1}, \cdots,  C''_{N_{C}}\}$ and $M_{C}$ is the total number of outgoing fluxes for the set of configurations $\{C'_{1}, ...,  C'_{M_{C}}\}$ as described in Fig. \ref{fig:multibalance}. 
At steady state,  for any system,   the  fluxes must balance:  $\sum_{C'} P(C) W(C \rightarrow C') = \sum_{C''} P( C'') W( C'' \rightarrow C)$. We have denoted $P(C)$ be the probability of the configuration $C$ and it can move to the other configuration $C'$ with a dynamical rate $W(C\rightarrow C')$.    For systems    that satisfy a MB, these  steady state configurations  break into  many  conditions of the form,  
\begin{eqnarray} 
\sum_{i=1}^{M_{C}} P(C) W(C \rightarrow C'_{i})
 = \sum_{i=1}^{N_{C}} P( C''_{i}) W( C''_{i} \rightarrow C).\label{eq:master_MB}
\end{eqnarray}
Eq. (\ref{eq:master_MB}) describes that for every configuration $C$, the incoming fluxes from a group of configurations $\{ C''_{1}, \cdots,  C''_{N_{C}}\}$, are balanced by outgoing fluxes to another uniquely identified group of configurations $\{C'_{1}, ...,  C'_{M_{C}}\}$. As a special case of MB condition, for $N_{C} = M_{C} = 1$, if   
$C''_{1} \neq C_{1}'$, Eq. (\ref{eq:master_MB}) reduces to Pairwise balance balance condition (PWB) and for the simplest case when $C''_{1} = C_{1}'$, it becomes the well known Detailed balance condition (DB) corresponds to the  equilibrium case.

\subsubsection{Balance Conditions for ZRP Model in two dimensions:}
To solve ZRP in two dimensions (see Fig. \ref{fig:2Dzrp}) when the hop rates in all four directions are different, we employ MB condition.
\begin{enumerate}
\item A PWB condition, where, the flux, generated, by a particle hopping from site $(i,j)$ of a configuration $C \equiv \{ \cdots, n_{i,j+1}, \cdots , n_{i-1,j}, n_{i,j}, n_{i+1,j}, \cdots , n_{i,j-1}, \cdots \}$ to its up nearest neighbour ({\it i.e.} site $(i,j+1)$), can be balanced with the flux due to hopping of a particle from site $(i,j-1)$ of another configuration $C' \equiv \{ \cdots, n_{i,j+1}, \cdots , n_{i-1,j}, n_{i,j}-1, n_{i+1,j}, \cdots , n_{i,j-1}+1, \cdots \}$ to site $(i,j)$. At steady state, PWB condition is satisfied similarly as ZRP model \cite{Evans_Hanney}, if 
\begin{equation}
 u_{u}(n_{i,j}) = \frac{f(n_{i,j}-1)}{f(n_{i,j})}.
\end{equation}
\item A MB condition, where, for a configuration $C \equiv \{ \cdots, n_{i,j+1}, \cdots , n_{i-1,j}, n_{i,j}, n_{i+1,j},\\ \cdots , n_{i,j-1}, \cdots \}$, the fluxes generated by a particle hopping to its right and left nearest neighbours from site $(i,j)$, can be balanced with the flux obtained by hopping of a particle from site $(i,j+1)$ of another configuration $C'' \equiv \{ \cdots, n_{i,j+1}+1, \cdots , n_{i-1,j}, n_{i,j}-1, n_{i+1,j}, \cdots , n_{i,j-1}, \cdots \}$ to its down nearest neighbour ({\it i.e.} site $(i,j)$). The flux balance scheme in Eq. (\ref{eq:master_MB})  gives the following equation
\begin{eqnarray}
\fl u_{d}(n_{i,j+1}+1) P(\cdots , n_{i,j+1}+1, \cdots n_{i,j} -1, \cdots )   =\left[~u_{r} (n_{i,j}) + u_{l} (n_{i,j})~\right] P(\{n_{i,j}\}). \label{eq:2ZRP_MB_rate}
\end{eqnarray}
\end{enumerate}
Using MB condition one can show that an exact steady state solution is possible and FSS as in Eq. (\ref{eq:FSS_P}) can be obtained  only when the hop rates satisfy
\begin{eqnarray} \label{eq:FSS_MB}
 u_{r}(n) + u_{l}(n) = u_{u}(n) = u_{d}(n)= u(n) = \frac{f(n-1)}{f(n)}.
\end{eqnarray}
As an example,  for this particle hopping ZRP model in two dimensions let us define the following rate functions  
\begin{eqnarray}
 u_{r}(n) = a ~~{\rm and}~~ u_{l}(n) =1-a~~ & {\rm if}~~ n = 1 \label{eq:ex_u_n1},\\
 u_{r}(n) = e^{\varepsilon /2} ~~{\rm and}~~ u_{l}(n) =e^{- \varepsilon /2}~~ & {\rm if}~~ n > 1, \label{eq:ex_u_n}
 \end{eqnarray}
where the model parameters $a < 1/2$ and $\varepsilon$ is the potential-bias  that is taken to be positive.
Using the rates in Eqs. (\ref{eq:ex_u_n1}) and  (\ref{eq:ex_u_n}) we can obtain the steady state weight following Eq. (\ref{eq:FSS_weight_steadty})
\begin{equation}
\label{eq:ex_steady_state}
f(n)=\cases{1&for $n = 0,1$\\
\delta^{n-1}&for $n>1$\\} ~; ~~\delta = ( 2\cosh(\varepsilon /2) )^{-1}.
\end{equation}

\subsubsection{Negative differential mobility in two dimensional ZRP}

In this section we would like to discuss ZRP in two dimensions with specific choices of rates in Eqs. (\ref{eq:ex_u_n1}) and (\ref{eq:ex_u_n}) give rise to  negative differential response \cite{NDM_PKM} of the particles. Following the local detailed balance condition, we can define the driving fields or bias in terms of the asymmetric rate functions as $E_{n_{i}, n_{i+1}} = \ln \frac{u_{r}(n_{i})}{u_{l}(n_{i+1} +1)}$  acting on bonds with local configurations $(n_i, n_{i+1})$ for the dynamics
\begin{equation} \label{eq:NDM_dynamics}
\{n_{i}, n_{i+1} \} \mathop{\rightleftharpoons}^{u_{r}(n_{i})}_{u_{l}(n_{i+1} +1)}
\{n_{i}-1, n_{i+1}+1 \}.
\end{equation}
For the set of specific rate functions in  Eqs. (\ref{eq:ex_u_n1}) and (\ref{eq:ex_u_n}) we can calculate $E_{n_{i}, n_{i+1}}$. The value $n_{i} = 0$ is excluded as in this case $(n_{i}-1)$ becomes negative. 
\begin{equation}
\label{eq:NDM_Ematrix}
E_{n_{i},n_{i+1}}=\cases{\ln (\frac{a}{1-a})&for $n_{i} = 1, ~ n_{i+1} =0$\\
\ln (a) + \varepsilon / 2 &for $n_{i} = 1, ~ n_{i+1} > 0$\\
- \ln(1-a) + \varepsilon / 2 &for $n_{i} > 1, ~ n_{i+1} = 0$\\
~\varepsilon &for $n_{i} > 1, ~ n_{i+1} > 0$}
\end{equation}
$E_{n_{i},n_{i+1}}$ increases linearly in the positive direction with the increase of the bias parameter $\varepsilon$ for all $n_{i}$ and $n_{i+1}$. We can express the grand canonical partition function following Eq. (\ref{eq:FSS_Q}) as  $Z_{L}(z) = \sum_{N=0}^{\infty} z^{N} Q_{L,N} = [F(z)]^{L}$ with 
\begin{equation}
 F(z) = \sum_{n=0}^{\infty} z^{n} f(n) = 1+\frac{z}{1-\delta z}
\end{equation}
where the fugacity $z$ controls the particle density through the relation $\rho(z)= z F'(z)/F(z) = z [1+z-2z \delta +z^{2} \delta (\delta - 1)]^{-1}$. Finally the current is\begin{eqnarray}\label{eq:NDM_current}
\fl J =\frac{1}{F(z)} \sum_{n=1}^{\infty} [~u_{r}(n)-u_{l}(n)~] z^{n} f(n)  = \frac{1}{F(z)} [~(2a-1)z + 2 \sinh(\varepsilon /2) \frac{z^{2} \delta}{1-\delta z}~].
\end{eqnarray}
\begin{figure}[h]
\vspace*{.4 cm}
\centering
\includegraphics[height=5.0 cm]{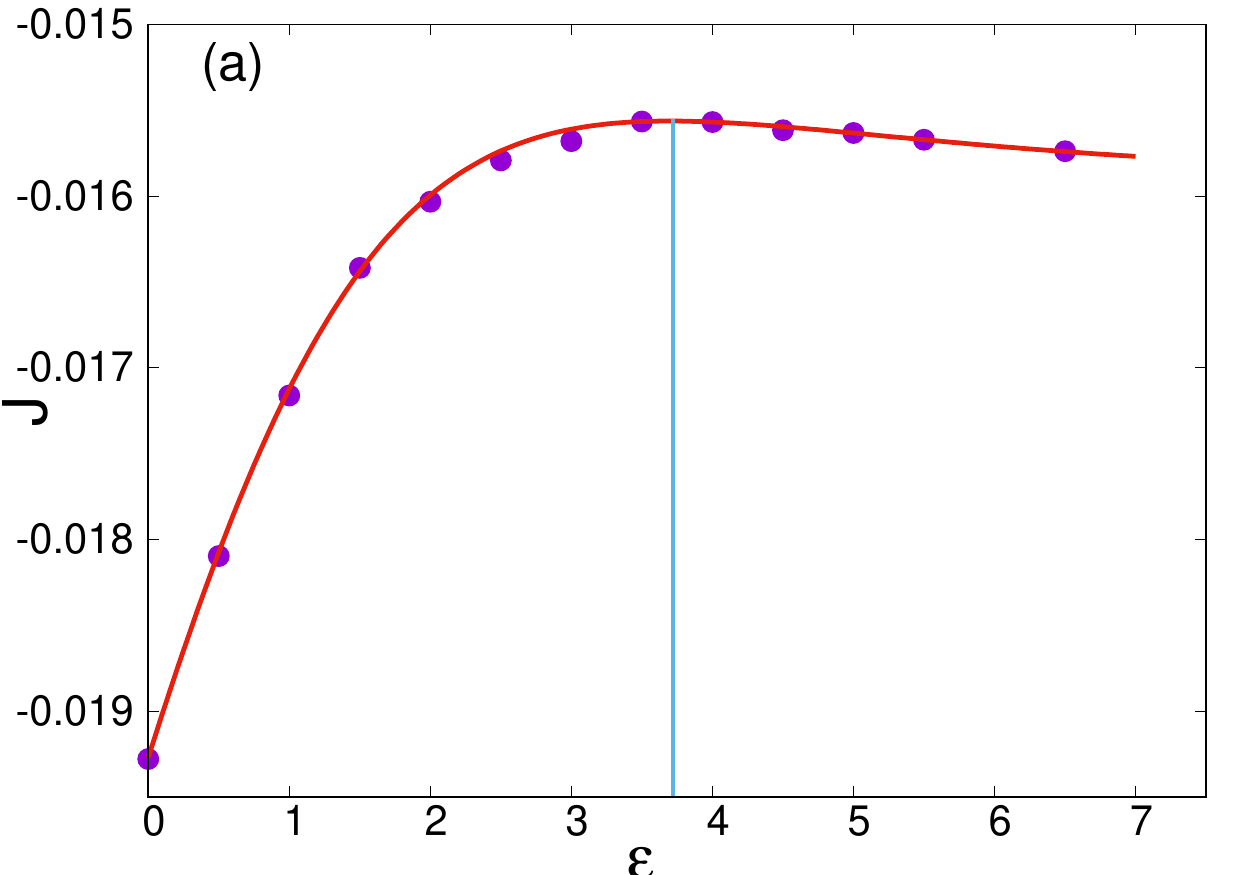} \hspace{.2 cm}\includegraphics[height=4.98 cm]{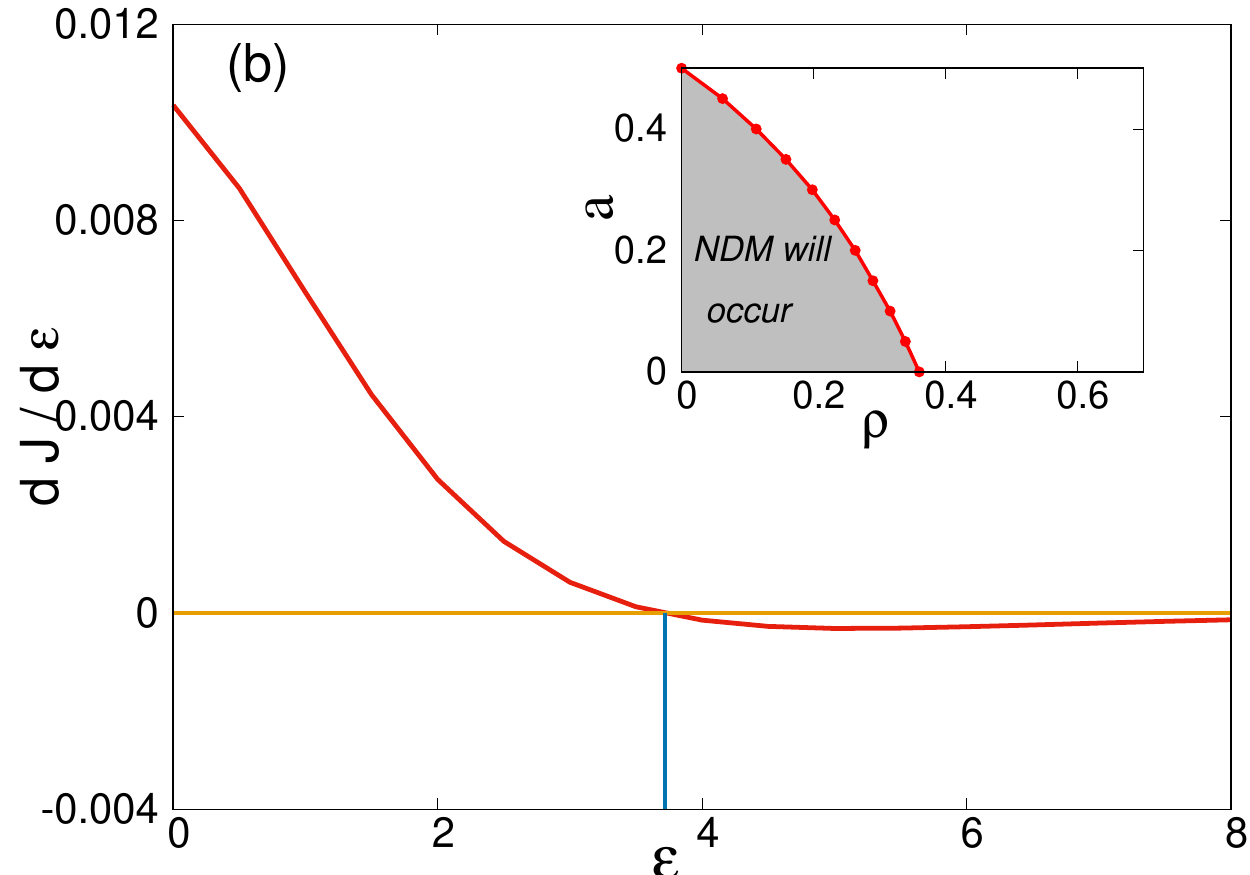}
\caption{(a) NDM in 2D ZRP, current ($J$) as a function of the bias ($\varepsilon$) measured for $\rho = 0.15$, $a = 0.2$, NDM occurs at $\varepsilon \geq 3.722$.  The 
results from simulation (points) with system size   $100 \times 100$ are compared with the exact result (line) according to Eq. (\ref{eq:NDM_current}), (b) $\frac{dJ}{d \varepsilon}$ as a function of $\varepsilon$ for $\rho = 0.15$, $a = 0.2$ using Eq. (\ref{eq:NDM_current}).  $\frac{dJ}{d \varepsilon}$ becomes negative for  $\varepsilon \geq 3.722$. Inset shows the phase diagram in $\rho$ - $a$ plane, NDM occurs in the  shaded region.}
 \label{fig:NDM}
\end{figure}
To understand the behaviour of current $J$ in Eq. (\ref{eq:NDM_current}), we did a Monte Carlo simulation for a fixed particle density $\rho = 0.15$ and $a = 0.2$. Fig. \ref{fig:NDM}(a) shows the particle current $J$ with the bias parameter $\varepsilon$ for $\rho = 0.15$ and $a = 0.2$. For small $\varepsilon$, current $J$ increases as the parameter $\varepsilon$  is increased. Beyond a certain value of $\varepsilon$ $(\varepsilon = 3.722)$ where $J$ reaches to its maximum value, it decreases with  $\varepsilon$ and NDM is observed as soon as $\varepsilon \geq 3.722$. It is evident from Fig. \ref{fig:NDM}(a) that the gradient of current $J$, $\frac{dJ}{d\varepsilon}$, decreases with $\varepsilon$ and becomes negative for $\varepsilon \geq 3.722$, which is shown in Fig. \ref{fig:NDM}(b). Current $J$ obtained in Eq. (\ref{eq:NDM_current}) may or may not exhibit NDM for every $\rho$ and $a$. To explore the possibility of NDM in this system, we have shown the region where NDM occurs in $\rho$ - $a$ phase plane in the inset of Fig. \ref{fig:NDM}(b).

\subsection{ZRP in three dimensions with asymmetric rates}

\begin{figure}[h]
\vspace*{.5 cm}
\begin{center}
\includegraphics[scale = 0.25]{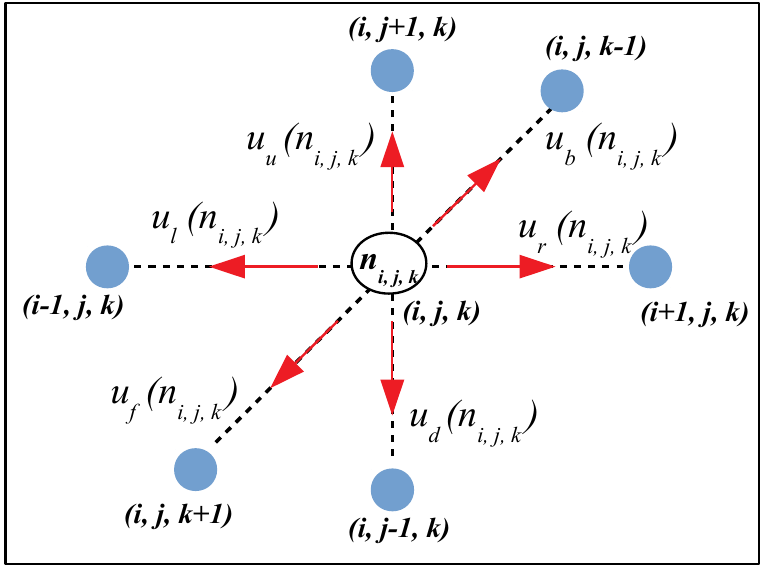}
\end{center}
\caption{  Asymmetric ZRP in three dimensions, where a particle from site $(i,j,k)$ can hop to its right, left, up and down, front and back nearest neighbours with rates $u_{r}(n_{i,j,k})$, $u_{l}(n_{i,j,k})$, $u_{u}(n_{i,j,k})$, $u_{d}(n_{i,j,k})$, $u_{f}(n_{i,j,k})$ and $u_{b}(n_{i,j,k})$ respectively, $n_{i,j,k}$ is number of particles at this site $(i,j,k)$.}
 \label{fig:3D_zrp}
\end{figure}

We consider a  periodic ZRP lattice in three dimensions of size ($L \times L \times L $) (see  Fig. \ref{fig:3D_zrp}). Each site,  represented by $(i, j, k)$ can be either vacant or it can be occupied by one or more particles denoted by $n_{i,j,k} \leq N$ $\left(N = \sum_{i,j,k=1}^{L} n_{i,j,k} \right )$. From a randomly chosen site $(i, j, k)$, a particle can hop to nearest neighbours (up, down, right, left, frnt and back)  with rates $u_{u}(n_{i,j,k})$, $u_{d}(n_{i,j,k})$, $u_{r}(n_{i,j,k})$, $u_{l}(n_{i,j,k})$, $u_{f}(n_{i,j,k})$ and  $u_{b}(n_{i,j,k})$.  
The steady state probability can be defined as 
\begin{equation} \label{eq:3DZRP_Pn}
 P(\{ n_{i,j,k}) \propto \prod_{i=1, j=1, k =1}^{L} f (n_{i,j,k})~ \delta \left(\sum_{i=1,j=1,k =1}^{L} n_{i,j,k} -N \right).
\end{equation}
When all rates are different, no general solution is available. One can obtain the FSS as in Eq. (\ref{eq:FSS_P}) for this model using  MB condition when the hop rates satisfy the condition 
\begin{equation} \label{eq:3ZRP_con_rate}
\fl [~u_{r} (n) + u_{l} (n) + u_{f} (n) + u_{b} (n)~] = u_{u}(n_{i,j,k})= u_{d}(n_{i,j,k}) =u (n) = \frac{f (n-1)}{f (n)}
\end{equation}
and the steady state weight is defined as  
$f(n) = \prod_{\nu =1}^{n} u(\nu)^{-1}$.  
As an example, let us define a simple choices of hop rates
\begin{eqnarray}
u(n) = (1+ \frac{1}{n})^{b} \label{eq:3ZRP_un},\\
 u_{r}(n) = [u(n) + \frac{a}{2}]/4 ~~{\rm and} ~~
 u_{l}(n)=[u(n) - \frac{a}{2}]/4 \label{eq:3ZRP_rl},\\
 u_{f}(n)= [u(n) + \frac{a}{3}]/4 ~~{\rm and} ~~
 u_{b}(n)= [u(n) - \frac{a}{3}]/4. \label{eq:3ZRP_fb}
\end{eqnarray}
\begin{figure}[h]
\vspace*{.5 cm}
\centering`
\includegraphics[height=5.0 cm]{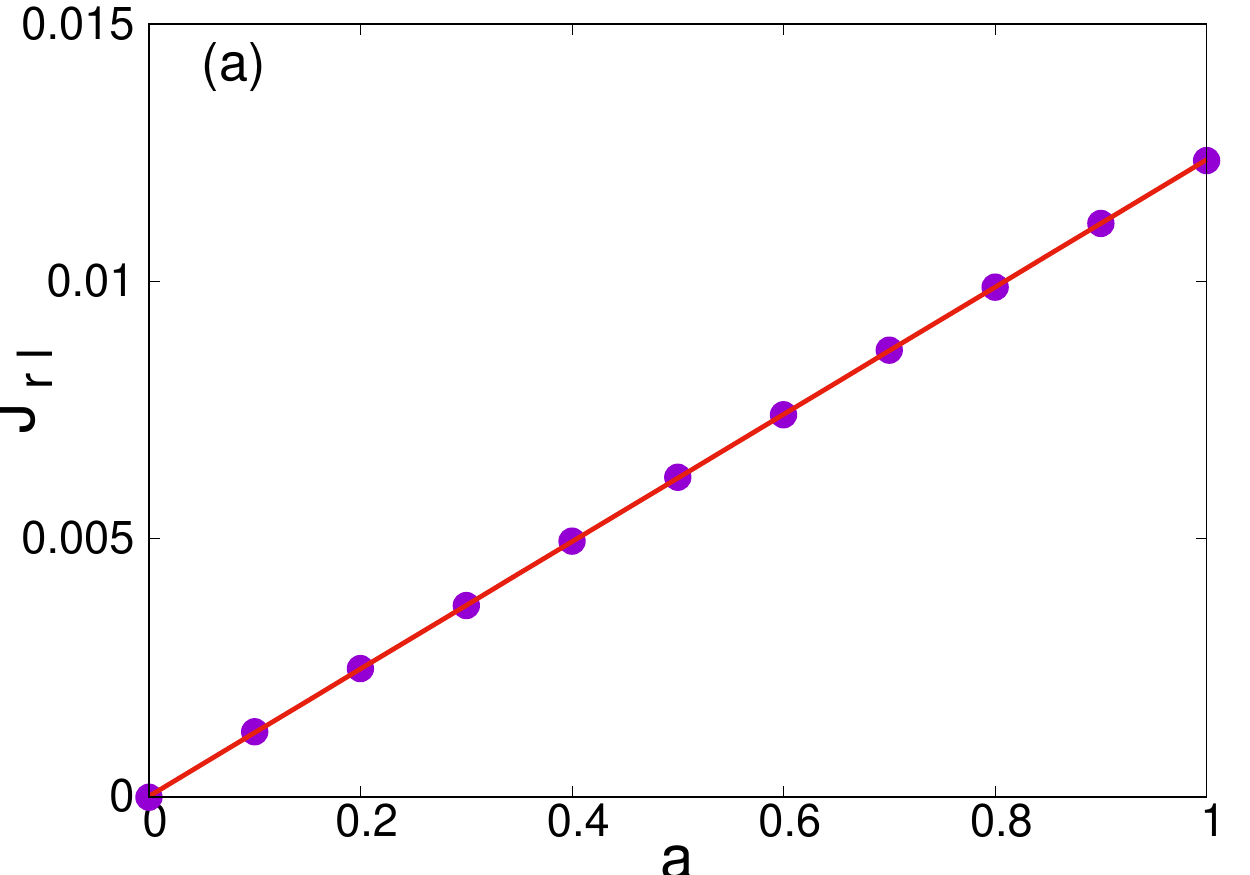} \hspace{.2 cm}\includegraphics[height=5.0 cm]{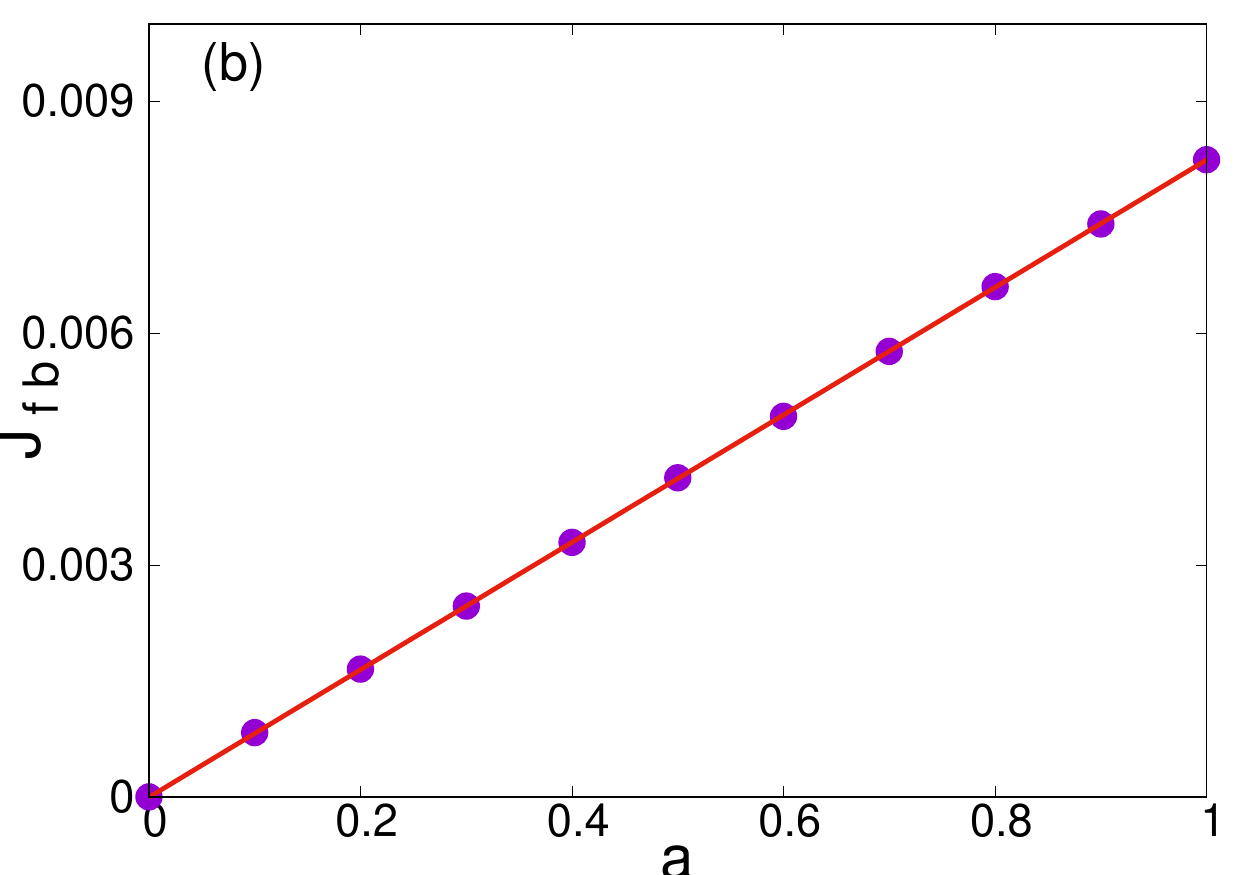}
\caption{Currents in three dimensional ZRP, (a) current $J_{rl}$ as a function of the parameter $a$ and  (b) current $J_{fb}$ as a function of the parameter $a$. The results from simulation (points) are compared with the exact results (line) according to Eqs. (\ref{eq:3ZRP_jrl}) and  (\ref{eq:3ZRP_jfb}). In both figures density $\rho$ = 0.5 and $b=1.1$} 
 \label{fig:3ZRP_curr}
\end{figure}
The grand canonical partition function can be expressed as $Z_{L}(z) = \sum_{N=0}^{\infty} z^{N} Q_{L,N} = [F(z)]^{L}$ with $F(z) = \sum_{n=0}^{\infty} z^{n} f(n) = (z)^{-1} Li_{b}(z)$ and $Li_{b}(z)$ is the Polylogarithm function defined by $Li_{b}(z) = \sum_{n=1}^{\infty} \frac{z^{n}}{n^{b}}$. Using our choices of rates in Eqs. (\ref{eq:3ZRP_un}) - (\ref{eq:3ZRP_fb}) we can calculate the currents $J_{rl}$ (right - left direction) and  $J_{fb}$ (front - back direction) as
\begin{eqnarray} 
 J_{rl} = \frac{1}{F(z)} \sum_{n=1}^{\infty} [u_{r}(n)-u_{l}(n)] f(n)z^{n} = 
 \frac{1}{F(z)} \frac{a}{4} (F(z) - 1 ), \label{eq:3ZRP_jrl}\\
 J_{fb} =\frac{1}{F(z)} \sum_{n=1}^{\infty} [u_{f}(n)-u_{b}(n)] f(n)z^{n} = \frac{1}{F(z)} \frac{a}{6} (F(z) - 1 ). \label{eq:3ZRP_jfb}
 \end{eqnarray}
In Fig. \ref{fig:3ZRP_curr}(a) and  Fig. \ref{fig:3ZRP_curr}(b) we have plotted the currents $J_{rl}$ and $J_{fb}$ with the parameter $a$ for particle density $\rho=0.5$ and $b=1.1$. As expected currents in both directions increase linearly with the parameter $a$.
\section{Asymmetric finite range process (FRP) with nearest neighbours and next nearest neighbours hopping}

\begin{figure}[h]
\vspace*{.5 cm}
\begin{center}
\includegraphics[scale = 0.3]{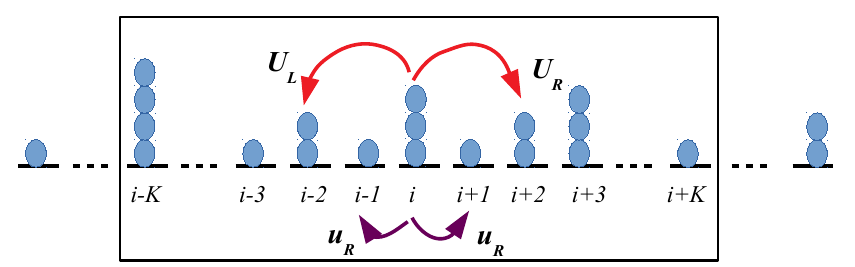}
\end{center}
\caption{ FRP in one dimension where a particle hop from a site $i$ to its left and right neighbours with rates $u_{L}(.)$  and $u_{R}(.)$ and left and right next nearest neighbours, with rates $U_{L}(.)$ and $U_{R}(.)$. All these rates  depends on occupation of site $i$ (here $n_{i} = 3$) and all its neighbours within a range $K$.}
\label{fig:MB_FRP_pic}
\end{figure}

We consider one dimensional  periodic lattice with sites labeled by $i = 1,2,\cdots ,L$ (see Fig. \ref{fig:MB_FRP_pic}). Each site $i$ has a non negative integer variable $n_{i}$, representing the number of particles at site $i$ (for a vacant site $n_{i} = 0$).
A particle from any randomly chosen site $i$; can hop to its right nearest neighbour with rate $u_{R}(.)$ and left  nearest neighbour with rate $u_{L}(.)$ and as well as to the next nearest neighbours with rates $U_{R}(.)$ for right, $U_{L}(.)$ for left. All these rates depend on the number of particles at all the sites within a range $K$ w.r.t the departure site. This finite range process (FRP), with nearest neighbours hopping has been studied earlier \cite{FRP, asymm_FRP}.
Like the FRP model discussed in \cite{FRP}, we  have considered the steady state probability
\begin{equation} \label{eq:MB_FRP_prob}
 P(\{n_{i}\}) = \frac{1}{Q_{L,N}} \prod_{i=1}^{L} 
 g(n_{i}, n_{i+1}, \cdots n_{i+K})
 ~\delta \left(\sum_{i} n_{i} - N\right)
\end{equation}
where, $Q_{L,N}$ is the canonical partition function defined as
\begin{equation}
 Q_{L,N} = \sum_{\{n_{i}\}} \prod_{i=1}^{L} 
 g(n_{i}, n_{i+1}, \cdots n_{i+K})
 ~\delta \left(\sum_{i} n_{i} - N\right).
\end{equation}
$N$ is the total number of particles and $\rho = \frac{N}{L}$ is conserved by the dynamics. 

\subsection{Balance conditions for asymmetric FRP}

We will try to get the steady states of this model for the asymmetric rates.   Consider the balance conditions as
\begin{enumerate}
\item A PWB condition, where the flux generated due to a particle hopping from site $i$ of a configuration $C \equiv \{ \cdots , n_{i-1}, n_{i}, n_{i+1}, \cdots \}$ to site $(i+1)$, can be balanced with the flux obtained by a particle hopping from site $(i-1)$ of another configuration $C' \equiv \{ \cdots , n_{i-1}+1, n_{i}-1, n_{i+1}, \cdots \}$ to site $i$.
\begin{eqnarray}
\fl u_{R}(n_{i-K}, \cdots , n_{i}, \cdots n_{i+K}) P(\{n_{i}\}) = &
 u_{R}(n_{i-K-1}, \cdots ,n_{i-1}+1, n_{i}-1, \cdots n_{i+K-1}) \cr & \times P(\cdots , n_{i-1}+1, n_{i}-1, n_{i+1}, \cdots ).\label{eq:MB_FRP_PWB1}
 \end{eqnarray}
Following Eq.  (\ref{eq:MB_FRP_prob}), one can check that PWB condition as in  Eq. (\ref{eq:MB_FRP_PWB1}) will be satisfied if the hop rate at site $i$ satisfies the condition as like FRP model \cite{FRP} 
 \begin{equation} \label{eq:MB_FRP_uRK}
  u_{R}(n_{i-K}, \cdots , n_{i}, \cdots , n_{i+K)}) = \prod_{k=1}^{K}
  \frac{g(\widetilde n_{i-K+k}, \widetilde n_{i-K+1+k},\cdots , \widetilde n_{i+k} )}{g(n_{i-K+k},n_{i-K+1+k}, \cdots , n_{i+k} )}
  \end{equation}
where $\widetilde n_{j} = n_{j} - \delta_{ji}$\\
\item A MB condition, where fluxes generated due to a particle hopping from  site $i$ of a configuration $C \equiv \{ \cdots , n_{i-1}, n_{i}, n_{i+1}, \cdots \}$ to the sites $(i+2)$ and $(i-2)$, can be balanced with the flux obtained by a particle hopping from site $(i+1)$ of another configuration $C'' \equiv \{ \cdots , n_{i-1}, n_{i}-1, n_{i+1}+1, \cdots \}$ to its left nearest neighbour site $i$.
The flux balance scheme in Eq. (\ref{eq:master_MB}) gives the following equation
\begin{eqnarray}
\fl [~U_{R}(n_{i-K}, \cdots , n_{i}, \cdots n_{i+K}) + U_{L}(n_{i-K}, \cdots , n_{i}, \cdots n_{i+K})~] ~P(\{n_{i}\}) \cr
\fl = u_{L}(n_{i-K+1}, \cdots ,n_{i-1}, n_{i}-1, n_{i+1}+1, \cdots n_{i+K+1})~ P(\cdots , n_{i-1}, n_{i}-1, n_{i+1}+1, \cdots ).\cr\label{eq:MB_FRP_MBK}
 \end{eqnarray}
\end{enumerate}
MB condition as in Eq. (\ref{eq:MB_FRP_MBK}) for this asymmetric FRP model will be satisfied and one can obtain a cluster-factorized  form of steady state (CFSS) as in Eq. (\ref{eq:MB_FRP_prob}), when $u_{R}(.) =u_{L}(.)=  u(.) $ and the hop rates satisfy the condition
\begin{eqnarray}
 U_{R}(n_{i-K}, \cdots , n_{i}, \cdots n_{i+K}) +U_{L}(n_{i-K}, \cdots , n_{i}, \cdots n_{i+K}) \cr = u(n_{i-K}, \cdots , n_{i}, \cdots n_{i+K})  = \prod_{k=1}^{K}
  \frac{g(\widetilde n_{i-K+k}, \widetilde n_{i-K+1+k},\cdots , \widetilde n_{i+k} )}{g(n_{i-K+k},n_{i-K+1+k}, \cdots , n_{i+k} )}\label{eq:MB_FRP_cond}
 \end{eqnarray}
where $\widetilde n_{j} = n_{j} - \delta_{ji}$

\subsection{Conditions for PFSS ($K=1$)}

For $K=1$, we can obtain the steady states in pair-factorized form using MB, if the hop rates satisfy, following Eq. (\ref{eq:MB_FRP_cond}),
\begin{eqnarray}
\fl [U_{R}(n_{i-1}, n_{i}, n_{i+1}) 
 + U_{L}(n_{i-1}, n_{i}, n_{i+1})] = 
 u(n_{i-1}, n_{i}, n_{i+1}) = &
 \frac{g(n_{i-1}, n_{i}-1)}{g(n_{i-1}, n_{i})}& \cr & \times ~\frac{g(n_{i}-1, n_{i+1})}{g(n_{i}, n_{i+1})}. &
\end{eqnarray}
Let us consider that the weight function $g(n_{i},n_{i+1})$ can be written as the inner product of two 2-dimensional vectors \cite{FRP}
\begin{equation}\label{eq:MB_FRP_gmn}
 g(n_{i},n_{i+1}) = \langle \alpha (n_{i}) | \beta (n_{i+1}) \rangle.
\end{equation}
In grand canonical ensemble where the fugacity $z$ controls the density $\rho$, the partition sum can be written as 
$Z_{L}(z)= \sum_{N=0}^{\infty} Q_{L,N}z^{N}
= \Tr[T(z)^{L}]$ with
\begin{equation}\label{eq:MB_FRP_T}
 T(z) = \sum_{n=0}^{\infty} z^{n} |\beta (n) \rangle \langle \alpha (n)|.
\end{equation}
Now, for a particular choice of the steady state weight $g(n_{i}, n_{i+1})$, one can construct the transfer matrix $T(z)$ to calculate the density of the system  $\rho = z \left(\frac{Z_{L}'(z)}{Z_{L}(z)}\right)$ and the correlation function by transfer matrix method in pair-factorized steady state \cite{FRP}.

\section{Two species Finite Range Process (FRP)}

\subsection{The Model}

\begin{figure}[h]
\vspace*{.5 cm}
\centering
\includegraphics[scale = 0.3]{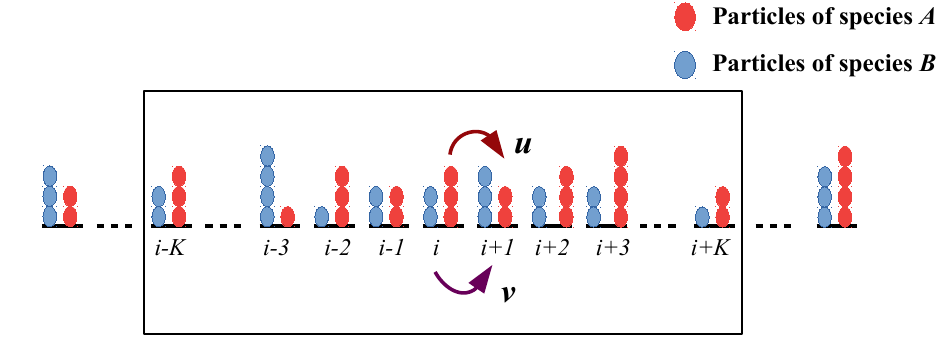}
\caption{Two species FRP in one dimension  where a particle of species $A$ hop from a site $i$ to its right neighbour with rate $u(.)$ and particle of species $B$ hops with rate $v(.)$. All these rates  depends on occupation of site $i$ (here $n_{i} = 3$, $m_{i} = 2$) and all its neighbours within a range $K$. }
 \label{fig:two_frp_model}
\end{figure}

The model is defined on an one dimensional periodic lattice with sites labeled by $i = 1,2, \cdots , L$ (see Fig. \ref{fig:two_frp_model}). At each site $i$, there are $n_{i}$ particles of species $A$ (coloured red) and $m_{i}$ particles of species $B$ (cloured blue). Total number of particle $A$ is $N$ and that of particle $B$ is $M$. A particle of any species, from any randomly chosen site $i$ can hop to its right nearest neighbour with a rate $u(.)$ for species $A$ and $v(.)$ for species $B$. These two rates  depend on the number of particles at all the sites which are within a range $K$ w.r.t the departure site like the FRP model \cite{FRP}. 
Dynamics of this model can be described as for species $A$
\begin{eqnarray}
\fl (\cdots , n_{i}, n_{i+1}, \cdots; \cdots, m_{i}, m_{i+1}, \cdots) 
 \to (\cdots,  n_{i}-1, n_{i+1}+1, \cdots; \cdots ,  m_{i}, m_{i+1}, \cdots) \cr  \fl {\rm with~rate}~~ u(n_{i-K}, \cdots , n_{i}, \cdots , n_{i+K}, m_{i-K}, \cdots , m_{i}, \cdots , m_{i+K})
\end{eqnarray}
and the dynamics of species $B$
\begin{eqnarray}
\fl (\cdots , n_{i}, n_{i+1}, \cdots; \cdots , m_{i}, m_{i+1}, \cdots) 
 \to (\cdots ,  n_{i}, n_{i+1}, \cdots; \cdots ,   m_{i}-1, m_{i+1}+1, \cdots) \cr \fl {\rm with~ rate}~~
 u(n_{i-K}, \cdots , n_{i}, \cdots , n_{i+K}, m_{i-K}, \cdots , m_{i}, \cdots , m_{i+K}).
\end{eqnarray}
For $K = 0$, this model is identical to two species zero range process  \cite{Evans_Hanney} with hop rate $u(.)$ for particles of species $A$, and $v(.)$ for particles of species $B$, an exactly solvable nonequilibrium model that evolves to a FSS
\begin{equation}
 \label{eq:MB_2ZRP_P}
 P(\{n_{i}\};\{m_{i}\})  \propto \prod_{i=1}^{L} 
 f(n_{i}, m_{i} )~
 \delta \left(\sum_{i} n_{i} - N\right)~ \delta \left(\sum_{i} m_{i} - M\right)
\end{equation}
with the steady state weight 
\begin{equation}
 f(n,m) = \prod_{i=1}^{n} [u(i,m)]^{-1} ~ \prod_{j=0}^{m} [v(0,j)]^{-1}.
\end{equation}

\subsubsection{Condition for cluster-factorized steady state (CFSS)}

We can express the steady state  probability $P(\{n_{i}\};\{m_{i}\})$ in a cluster factorized form as 
\begin{eqnarray} \label{eq:MB_2FRP_P}
 P(\{n_{i}\};\{m_{i}\})  = & \frac{1}{Q_{L,N,M}} \prod_{i=1}^{L} 
 g(n_{i}, n_{i+1}, \cdots n_{i+K},m_{i}, m_{i+1}, \cdots m_{i+K} )& \cr &
 \times \delta \left(\sum_{i} n_{i} - N\right) ~\delta \left(\sum_{i} m_{i} - M\right)&
\end{eqnarray}
where $Q_{L,N,M}$ is the canonical partition function defined as
\begin{eqnarray} \label{eq:MB_2FRP_Q}
\fl Q_{L,N,M} =\sum_{\{n_{i},m_{i}\}} \prod_{i=1}^{L} 
 g(n_{i}, \cdots n_{i+K},m_{i}, \cdots m_{i+K} )
~\delta \left(\sum_{i} n_{i} - N\right) ~\delta \left(\sum_{i} m_{i} - M\right).
\end{eqnarray}
$N$ and $M$ are total number of particles of species $A$ and $B$. $\rho_{A} = \frac{N}{L}$ and $\rho_{B} = \frac{M}{L}$ are conserved by the dynamics.
We can write the Master equation of this two species FRP model to find the steady state condition
\begin{eqnarray} 
\fl \frac{d}{dt} P(\{n_{i}\};\{m_{i}\}) =& 
 \sum_{i=1}^{L} [-u(n_{i-K}, \cdots , n_{i},\cdots, n_{i+k}, \cdots,m_{i-k},\cdots, m_{i}, \cdots, m_{i+K})  P(\{n_{i}\};\{m_{i}\}) & \cr
 & +u(n_{i-K}, \cdots , n_{i}-1, n_{i+1}+1,\cdots, n_{i+k}, \cdots, m_{i-k},\cdots,m_{i}, \cdots, m_{i+K} )& \cr 
  & \times P(\cdots , n_{i}-1, n_{i+1}+1, \cdots ;  \{m_{i}\} ) ] + & \cr
 & \sum_{i=1}^{L} [-v(n_{i-K}, \cdots , n_{i},\cdots, n_{i+k}, \cdots,m_{i-k},\cdots, m_{i}, \cdots, m_{i+K}) P(\{n_{i}\};\{m_{i}\})& \cr
 &+ v(n_{i-K}, \cdots, n_{i},  \cdots,n_{i+k}, \cdots, m_{i-k}, \cdots, m_{i}-1, m_{i+1}+1, \cdots, m_{i+K} )& \cr
  & \times P( \{n_{i}\};  \cdots ,  m_{i}-1, m_{i+1}+1, \cdots )].& \label{eq:MB_2FRP_ME}
\end{eqnarray}
Eq. (\ref{eq:MB_2FRP_ME}) is true when the gain and loss terms due to the dynamics of species $A$  cancel independently of the gain and loss terms due to the dynamics of species $B$. We look to achieve this cancellation for each term $i$ in the sum separately. With this condition,  the cluster-factorized form of steady states (CFSS) as in Eq. (\ref{eq:MB_2FRP_P}) for this two species FRP model is indeed possible when the hop rates of species $A$ and $B$ satisfy the conditions\newpage
\begin{eqnarray} 
 u(n_{i-K}, \cdots , n_{i}, \cdots n_{i+K}
 ,m_{i-K}, \cdots , m_{i}, \cdots m_{i+K} ) \cr
 =
 \prod_{k=1}^{K}
  \frac{g(\widetilde n_{i-K+k}, \widetilde n_{i-K+1+k},\cdots , \widetilde n_{i+k}, m_{i-K+k}, m_{i-K+1+k},\cdots , m_{i+k}
  )}
  {g(n_{i-K+k},n_{i-K+1+k}, \cdots , n_{i+k},m_{i-K+k},m_{i-K+1+k}, \cdots , m_{i+k} )},
  \label{eq:MB_2FRP_rate_u}\\
v(n_{i-K}, \cdots , n_{i}, \cdots n_{i+K}
 ,m_{i-K}, \cdots , m_{i}, \cdots m_{i+K} )
 \cr =
 \prod_{k=1}^{K}
  \frac{g( n_{i-K+k}, n_{i-K+1+k},\cdots , n_{i+k}, \widetilde m_{i-K+k}, \widetilde m_{i-K+1+k},\cdots , \widetilde m_{i+k}
  )}
  {g(n_{i-K+k},n_{i-K+1+k}, \cdots , n_{i+k},m_{i-K+k},m_{i-K+1+k}, \cdots , m_{i+k} )}  \label{eq:MB_2FRP_rate_v}
\end{eqnarray}
where $\widetilde n_{j} = n_{j} - \delta_{ji}$ in Eq. (\ref{eq:MB_2FRP_rate_u}) and $\widetilde m_{j} = m_{j} - \delta_{ji}$ in Eq. (\ref{eq:MB_2FRP_rate_v}). 
These two rates are related by a constraint 
\begin{eqnarray}
\frac{u(n_{i-K}, \cdots , n_{i}, \cdots n_{i+K}
 ,m_{i-K}, \cdots , m_{i}, \cdots m_{i+K} )}{u(n_{i-K}, \cdots , n_{i}, \cdots n_{i+K}
 ,m_{i-K}, \cdots , m_{i}-1, \cdots m_{i+K} )} \cr = 
 \frac{v(n_{i-K}, \cdots , n_{i}, \cdots n_{i+K}
 ,m_{i-K}, \cdots , m_{i}, \cdots m_{i+K} )}{v(n_{i-K}, \cdots , n_{i}-1, \cdots n_{i+K}
 ,m_{i-K}, \cdots , m_{i}, \cdots m_{i+K} )}. \label{eq:MB_2FRP_uvcon}
\end{eqnarray}
As like ZRP model with several species of particles \cite{Evans_Hanney, Evans_Hanney_conden, Grosskinsky_ZRP} , it is possible to generalise  FRP to any number of species say $Q$. Although there are $Q$ rates, it is expected that the rates must be related by $Q-1$ conditions.

\subsection{Two species FRP model with directional asymmetry}

\begin{figure}[h]
\centering
\includegraphics[scale = 0.3]{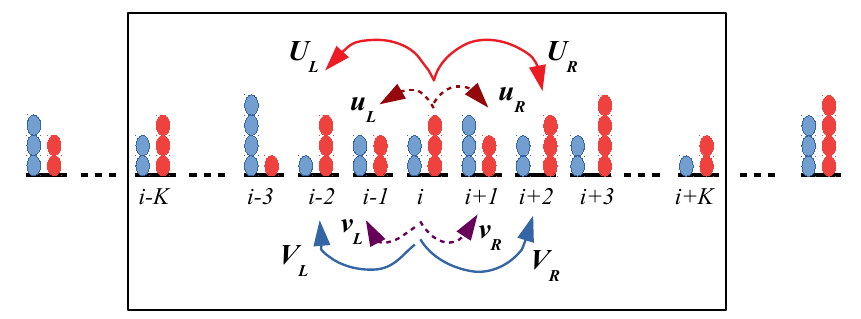}
\caption{Two species asymmetric FRP  model in one dimension. A particle of species A can hop to its right and left  nearest neighbours with rates $u_{R}(.)$ and $u_{L}(.)$, to right and left next nearest neighbours with rates $U_{R}(.)$ and $U_{L}(.)$. Similarly, particle of species $B$, can hop to its right and left nearest neighbours with rates $v_{R}(.)$ and $v_{L}(.)$, to right and left next nearest neighbours with rates $V_{R}(.)$ and $V_{L}(.)$.}
 \label{fig:MB_frp_model}
\end{figure}

We can add a directional asymmetry in  two species FRP model (see Fig. \ref{fig:MB_frp_model}), by adding conditions,  from a  randomly chosen site $i$, the particle of species $A$, can hop to its right and left nearest neighbours with rates $u_{R}(.)$ and $u_{L}(.)$, it can hop to right and left next nearest neighbours with rates $U_{R}(.)$ and $U_{L}(.)$. Similarly, particle of species $B$, can hop to its right and left nearest neighbours with rates $v_{R}(.)$ and $v_{L}(.)$, to right and left next nearest neighbours with rates $V_{R}(.)$ and $V_{L}(.)$. All these rates  depend on the number  of particles at all the sites which are within a range $K$ w.r.t the departure site Fig. \ref{fig:MB_frp_model}.

\subsubsection{Balance conditions for two species FRP model with directional asymmetry}

For $K = 1$ (PFSS), we can express the steady state probability $P(\{n_{i}\};\{m_{i}\})$ in terms of the steady state weight following Eq. (\ref{eq:MB_2FRP_P})
\begin{eqnarray} 
\fl P(\{n_{i}\};\{m_{i}\}) = \frac{1}{Q_{L,N,M}} \prod_{i=1}^{L} 
 g(n_{i}, n_{i+1},m_{i}, m_{i+1} )
~\delta \left(\sum_{i} n_{i} - N\right) \delta \left(\sum_{i} m_{i} - M\right) \label{eq:MB_2FRP_P_PFSS}
\end{eqnarray}
where, the canonical partition function 
\begin{equation}\label{eq:MB_2FRP_QLN_PFSS}
\fl Q_{L,N,M} =\sum_{\{n_{i};m_{i}\}}\prod_{i=1}^{L} 
 g(n_{i}, n_{i+1},m_{i}, m_{i+1} )
~ \delta \left(\sum_{i} n_{i} - N\right) \delta \left(\sum_{i} m_{i} - M\right).
\end{equation}
Consider the balance conditions to obtain the steady state 
\begin{enumerate}
\item PWB conditions,  where\\
(a) Flux generated by hopping of a particle of species $A$ from site $i$ of a configuration $C \equiv (\cdots , n_{i-1}, n_{i}, n_{i+1}, \cdots, \cdots ,  m_{i-1}, m_{i}, m_{i+1}, \cdots)$ to  site $(i+1)$, can be balanced with the flux obtained by hopping of a particle of same species from site $(i-1)$ of another configuration $C' \equiv (\cdots , n_{i-1}+1, n_{i}-1, n_{i+1}, \cdots, \cdots ,  m_{i-1}, m_{i}, m_{i+1}, \cdots)$ to  site $i$.\\
(b) Similarly, flux generated by hopping of a particle of species $B$ from site $i$ of the configuration $C \equiv (\cdots , n_{i-1}, n_{i}, n_{i+1}, \cdots, \cdots ,  m_{i-1}, m_{i}, m_{i+1}, \cdots)$ to site $(i+1)$, can be balanced with the flux obtained by hopping of a particle of same species from site $(i-1)$ of another configuration $C'' \equiv (\cdots , n_{i-1}, n_{i}, n_{i+1}, \cdots, \cdots ,  m_{i-1}+1, m_{i}-1, m_{i+1}, \cdots)$ to site $i$. 
For these PWB conditions with similar argument like Eq.  (\ref{eq:MB_2FRP_ME}), we can calculate rates of species $A$ and $B$ respectively for $K=1$ as
\begin{eqnarray}
 \fl u_{R}(n_{i-1},n_{i},n_{i+1},m_{i-1},m_{i},m_{i+1})
 = \frac{g(n_{i-1},n_{i}-1, m_{i-1},m_{i})}{g(n_{i-1},n_{i}, m_{i-1},m_{i})} \frac{g(n_{i}-1,n_{i+1} m_{i},m_{i+1})}{g(n_{i},n_{i+1}, m_{i},m_{i+1})} \label{eq:MB_2FRP_u_PFSS},\cr\\
 \fl v_{R}(n_{i-1},n_{i},n_{i+1},m_{i-1},m_{i},m_{i+1})
 = \frac{g(n_{i-1},n_{i}, m_{i-1},m_{i}-1)}{g(n_{i-1},n_{i}, m_{i-1},m_{i})} \frac{g(n_{i},n_{i+1} m_{i}-1,m_{i+1})}{g(n_{i},n_{i+1}, m_{i},m_{i+1})}.\cr \label{eq:MB_2FRP_v_PFSS}
 \end{eqnarray}
\item MB conditions where\\
 (a) Fluxes generated for a particle of species $A$, hopping from site $i$, of configuration $C \equiv (\cdots , n_{i-1}, n_{i}, n_{i+1}, \cdots, \cdots ,  m_{i-1}, m_{i}, m_{i+1}, \cdots)$, to sites $(i+2)$ and $(i-2)$, can be balanced with the flux obtained by hopping of a particle of species $B$ from site $i$ of the configuration $C_{1}' \equiv \{ (\cdots , n_{i-1}, n_{i}, n_{i+1}, \cdots ,  m_{i-1}, m_{i}-1, m_{i+1}+1, \cdots)\}$ to site $(i-1)$.\\
(b) Fluxes generated for a particle of species $B$, hopping from site $i$, of configuration $C \equiv (\cdots , n_{i}, n_{i+1}, \cdots,  m_{i-1}, m_{i}, m_{i+1}, \cdots)$, to sites $(i+2)$ and $(i-2)$, can be balanced with the flux obtained by hopping of a particle of species $A$ from site $i$ of the configuration $C_{2}' \equiv \{ (\cdots , n_{i-1}, n_{i}-1, n_{i+1}+1, \cdots, \cdots ,  m_{i-1}, m_{i}, m_{i+1}, \cdots)\}$ to site $(i-1)$.
\end{enumerate}
One can verify that pair-factorized form of steady state (PFSS) as in Eq. (\ref{eq:MB_2FRP_P_PFSS}) can be obtained following these MB conditions when the rates $u_{R}(.) = u_{L}(.) = u(.)$, $v_{R}(.) = v_{L}(.) = v(.)$ and other hop rates of species $A$ and $B$ satisfy 
\begin{eqnarray} 
\fl \left[~U_{R}(n_{i-1},n_{i},n_{i+1},m_{i-1},m_{i},m_{i+1}) + U_{L}(n_{i-1},n_{i},n_{i+1},m_{i-1},m_{i},m_{i+1})~ \right] \cr
\fl = 
  v(n_{i-1},n_{i},n_{i+1},m_{i-1},m_{i},m_{i+1})
 = \frac{g(n_{i-1},n_{i}, m_{i-1},m_{i}-1)}{g(n_{i-1},n_{i}, m_{i-1},m_{i})} \frac{g(n_{i},n_{i+1} m_{i}-1,m_{i+1})}{g(n_{i},n_{i+1}, m_{i},m_{i+1})}\label{eq:MB_2FRP_con1_PFSS}
\end{eqnarray}
\begin{eqnarray} 
\fl \left[~V_{R}(n_{i-1},n_{i},n_{i+1},m_{i-1},m_{i},m_{i+1}) + V_{L}(n_{i-1},n_{i},n_{i+1},m_{i-1},m_{i},m_{i+1})~ \right] \cr  \fl = 
  u(n_{i-1},n_{i},n_{i+1},m_{i-1},m_{i},m_{i+1})
 = \frac{g(n_{i-1},n_{i}-1, m_{i-1},m_{i})}{g(n_{i-1},n_{i}, m_{i-1},m_{i})} \frac{g(n_{i}-1,n_{i+1} m_{i},m_{i+1})}{g(n_{i},n_{i+1}, m_{i},m_{i+1})}\label{eq:MB_2FRP_con2_PFSS}
\end{eqnarray}
with the rates $u(.)$ and $v(.)$ are related by a constraint 

\begin{equation}\label{eq:MB_2FRP_uvcon_PFSS}
\fl \frac{u(n_{i-1},n_{i},n_{i+1},m_{i-1},m_{i},m_{i+1})}{u(n_{i-1},n_{i},n_{i+1},m_{i-1},m_{i}-1,m_{i+1})} =
 \frac{v(n_{i-1},n_{i},n_{i+1},m_{i-1},m_{i},m_{i+1})}{v(n_{i-1},n_{i}-1,n_{i+1},m_{i-1},m_{i},m_{i+1})}.
\end{equation}

\subsubsection{Observable in two species FRP for K = 1 (PFSS)}

Let us consider that the weight function $g(n_{i},n_{i+1},m_{i},m_{i+1})$ in Eq. (\ref{eq:MB_2FRP_P_PFSS}) can be written by four 2-dimensional vectors as
\begin{equation}\label{eq:MB_2FRP_gfun_PFSS}
 g(n_{i},n_{i+1},m_{i},m_{i+1}) = \langle \alpha _{n_{i}}|\beta _{n_{i+1}}\rangle \langle \gamma _{m_{i}}|\delta _{m_{i+1}} \rangle.
\end{equation}
In grand canonical ensemble, the partition sum following Eq. (\ref{eq:MB_2FRP_QLN_PFSS}) becomes $Z_{L}(z,y) =\sum_{N=0}^{\infty} \sum_{M=0}^{\infty} z^{N} y^{M} Q_{L,N,M}  = \Tr[T_{1}(z)]^{L} ~ \Tr[T_{2}(y)]^{L}$
where, we now have two fugacities $z$ and $y$ that fix the particle densities 
of the species $A$ and $B$ with the transfer matrices 
\begin{equation}\label{eq:MB_2FRP_T1T2_PFSS}
 T_{1}(z) = \sum_{n=0}^{\infty} z^{n} |\beta (n)\rangle \langle\alpha (n)| ~~~{\rm and}~~~ 
 T_{2}(y) = \sum_{m=0}^{\infty} y^{m} |\delta (m)\rangle \langle\gamma (m)|.
\end{equation}
For an example, we consider the 2-dimensional representations as
\begin{eqnarray}
 \langle \alpha (n)| = (\frac{1}{(n+1)^{\nu}}, \frac{n+1}{(n+1)^{\nu}}) ~~{\rm and}~~ \langle \beta (n)| = (n+1,1) \\
\fl \langle \gamma (m)| = (\frac{m+1}{(m+1)^{\nu/2}}, \frac{1}{(m+1)^{\nu/2}}) ~~{\rm and}~~ \langle \delta (m)| = (\frac{1}{(m+1)^{\nu/2}},\frac{m+1}{(m+1)^{\nu/2}})
\end{eqnarray}
such that the steady state weight becomes 
\begin{equation} \label{eq:MB_2FRP_weightfn}
 g(n_{i},n_{i+1},m_{i},m_{i+1}) = \frac{(n_{i}+n_{i+1}+2)}{(n_{i}+1)^{\nu}} \frac{(m_{i}+m_{i+1}+2)}{(m_{i}+1)^{\nu/2} (m_{i+1}+1)^{\nu/2}}.
\end{equation}
In this case, one can calculate the desired hop rates of both species $A$ and species $B$ for which the PFSS with the weight function in Eq. (\ref{eq:MB_2FRP_weightfn}) is realized. The Transfer matrix $T_{1}(z)$ following Eq. (\ref{eq:MB_2FRP_T1T2_PFSS}) becomes 
\begin{equation}
T_{1}(z) = \frac{1}{z} \left(
\begin{array}{cc}
Li_{\nu -1}(z) & Li_{\nu -2}(z) \\
Li_{\nu}(z) & Li_{\nu -1}(z)
\end{array}
\right)
\end{equation}
and the transfer matrix $T_{2}(y)$ is just the transpose of the matrix $T_{1}(y)$
\begin{equation}\label{eq:MB_2FRP_T2Matrix}
 T_{2}(y) = [T_{1}(y)]^{T}.
\end{equation}
The eigenvalues of $T_{1}(z)$ and $T_{2}(y)$ are $\lambda _{\pm}(z)$ and 
$\chi_{\pm}(y)$ where
\begin{equation}
 \lambda _{\pm}(z) = \frac{1}{z} (Li_{\nu -1} (z) \pm \sqrt{Li_{\nu}(z) Li_{\nu-2} (z)}~) ~~{\rm and}~~ \chi_{\pm}(y) = \lambda _{\pm}(y).
\end{equation}
The partition function $Z_{L}(z,y)$ in the thermodynamic limit becomes $Z_{L}(z,y) = \left(\lambda_{+}(z)^{L} \chi_{+}(y)^{L} \right )$, as $\lambda_{+}(z)$ and $\chi_{+}(y)$ are the function of $z$ and $y$ only, we can write the density fugacity relation of species $A
$ as $\rho_{A}(z) = z \frac{\partial }{\partial z}\ln (\lambda_{+}(z))$ and for  species $B$ $\rho_{B}(y) = y \frac{\partial }{\partial y}\ln (\chi_{+}(y))$. 
\begin{figure}[h]
\vspace*{.5 cm}
\centering
\includegraphics[height=5.0 cm]{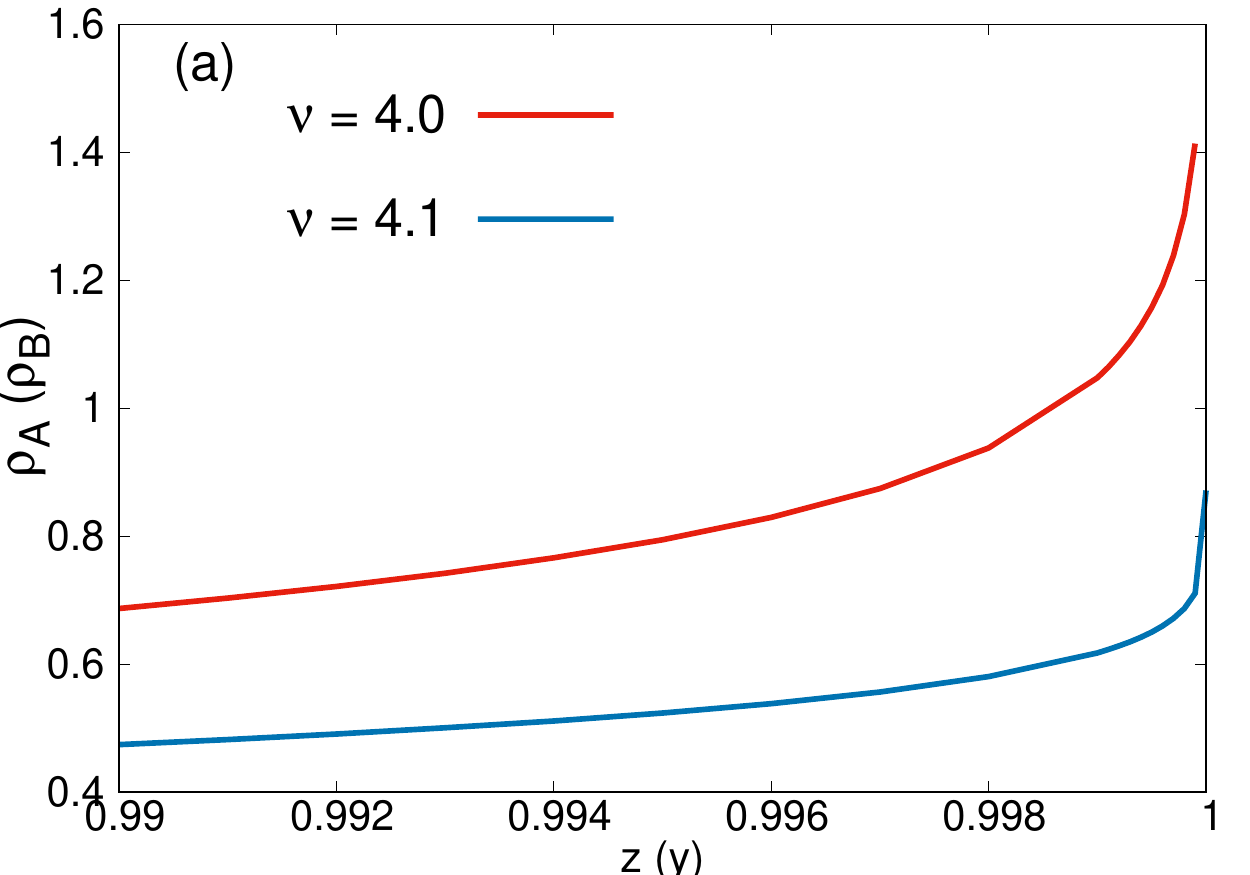} \hspace{.2 cm}\includegraphics[height=5.0 cm]{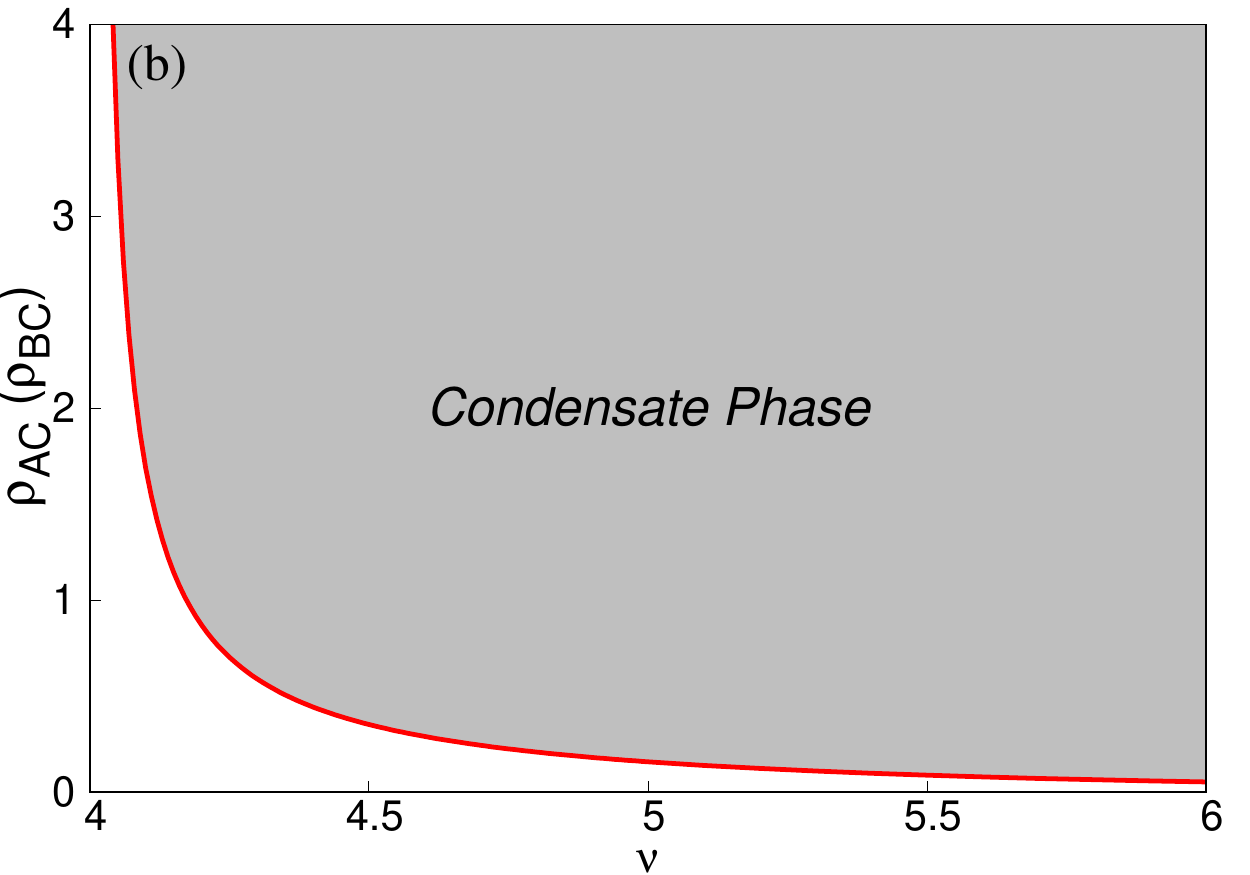}
\caption{Condensation in two species FRP, (a) curves are from exact calculation; $\rho_{A}$ $(\rho_{B})$ vs $z$ $(y)$ curve for $\nu = 4.0$ and $\nu = 4.1$, both $\rho_{A}$ and $\rho_{B}$  diverges for $\nu \leq 4.0$ and becomes finite for  $\nu > 4.0$, (b) Phase diagram of the condensation transition in $\rho_{A}$ $(\rho_{B})$ - $\nu$ plane}
 \label{fig:rho_A_B}
\end{figure}
The critical density of species $A$ be $\rho_{AC} = \lim_{z \rightarrow 1} \rho_{A}(z)$, and critical density of species $B$,  $\rho_{BC} = \lim_{y \rightarrow 1} \rho_{B}(y)$. It turns out that for both species $A$ and $B$, for $ \nu \leq 4$, the critical densities $\rho_{AC}$ and $\rho_{BC}$ diverge. For $\nu > 4$, as $\rho_{AC}$ and $\rho_{BC}$ give finite value as  shown in Fig. \ref{fig:rho_A_B} (a), we can say that we have a condensate when densities exceed the critical value for both species. Thus the model exhibits a phase transition between a fluid phase and a condensed phase where the excess particles condense onto a single site. The phase diagram of the condensation transition in $\rho_{A}$ $(\rho_{B})$ - $\nu$ plane is shown in Fig. \ref{fig:rho_A_B} (b). The critical line $\rho_{A}$ $(\rho_{B})$ separates the condensate phase from the fluid phase.

\section{Asymmetric hopping on a triangular lattice}

\begin{figure}[h]
\vspace*{.5 cm}
\centering
\includegraphics[scale = 0.35]{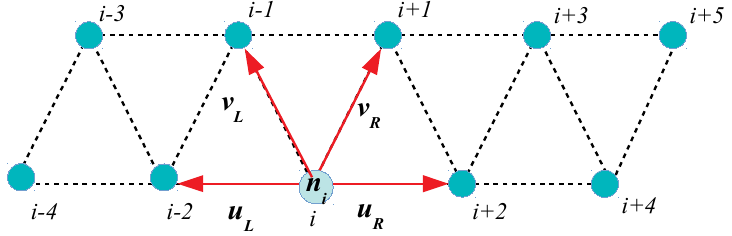}
\caption{Triangular lattice model, particle at site $i$ is represented by $n_{i}$. A particle from site $i$, can hop to sites $(i+1)$ and $(i-1)$  with rates $v_{R}(.)$ and $v_{L}(.)$ respectively and  can hop to sites $(i+2)$  and $(i-2)$  with rates $u_{R}(.)$ and $u_{L}(.)$ respectively. $n_{i}$ is the number of particles at site $i$.}
 \label{fig:triangular_lattice}
\end{figure}

We consider a periodic triangular lattice with sites labeled by $i = 1, 2, \cdots, L$ (see Fig. \ref{fig:triangular_lattice}). Each site $i$ has a non negative integer variable $n_{i}$, representing the number of particles at site $i$ ($n_{i} = 0$ if the site is vacant). A particle from any randomly chosen site $i$, can hop to sites $(i-1)$  and $(i+1)$  with rates $v_{L}(.)$ and $v_{R}(.)$ respectively and  can hop to the sites $(i-2)$  and $(i+2)$  with rates $u_{L}(.)$ and $u_{R}(.)$. Each of these rates depend on the number of particles of sites $(i-2,  i-1,  i,  i+1,  i+2)$. To obtain the steady state of this model for this asymmetric rate, we can consider the steady state probability as
\begin{equation} \label{eq:tri_Pn}
 P(\{ n_{i}\}) \propto \prod_{i=1}^{L} g(n_{i},n_{i+1}) h(n_{i},n_{i+2}) ~ \delta \left(\sum_{i=1}^{L} n_{i} - N \right).
\end{equation}
$N$ is the total number of particles and $\rho = \frac{N}{L}$ is conserved by the dynamics. We will try to obtain the steady state using the MB  condition.

\subsection{Balance conditions for triangular lattice}
\begin{enumerate}
\item A PWB condition where flux generated due to a particle hopping from site $i$ of the configuration $C \equiv \{ \cdots , n_{i-2}, n_{i-1}, n_{i}, n_{i+1}, n_{i+2}, \cdots \}$, to site $(i+2)$ is balanced with the flux obtained by a particle hopping from site $(i-2)$ of another configuration $C' \equiv \{ \cdots , n_{i-2}+1, n_{i-1}, n_{i}-1, n_{i+1}, n_{i+2}, \cdots \}$ to site $i$. The flux balance scheme described in Eq. (\ref{eq:master_MB}) gives the following condition
\begin{eqnarray}
u_{R} (n_{i-4},n_{i-3}, n_{i-2}+1, n_{i-1}, n_{i}-1) P(\cdots, n_{i-2}+1, n_{i-1}, n_{i}-1, \cdots ) \cr =u_{R} (n_{i-2},n_{i-1}, n_{i}, n_{i+1}, n_{i+2})  P(\{n_{i}\}) 
\end{eqnarray}
we can verify that the form of the steady state in pair factorized form as in 
Eq. (\ref{eq:tri_Pn}), is indeed possible when the hop rate $u(.)$ at site $i$ satisfies the following condition 
\begin{eqnarray}
 u_{R} (n_{i-2},n_{i-1}, n_{i}, n_{i+1}, n_{i+2}) &=  \frac{g(n_{i-1}, n_{i}-1)}{g(n_{i-1}, n_{i})}
\frac{g(n_{i}-1, n_{i+1})}{g(n_{i}, n_{i+1})}&\cr& \times 
 \frac{h(n_{i-2}, n_{i}-1)}{h(n_{i-2}, n_{i})}
\frac{h(n_{i}-1, n_{i+2})}{h(n_{i}, n_{i+2})}.& \label{eq:tri_uR2}
\end{eqnarray}
\item A MB condition where fluxes generated due to the particle hopping from site $i$ of the configuration $C \equiv \{ \cdots , n_{i-2}, n_{i-1}, n_{i}, n_{i+1}, n_{i+2}, \cdots \}$ to sites $(i-1)$ and $(i+1)$  are balanced by the flux obtained by hopping of a particle from site $(i+2)$  of another configuration $C'' \equiv \{ \cdots , n_{i}-1, n_{i+1}, n_{i+2}+1, \cdots \}$ to site $i$. The flux cancellation scheme in  Eq. (\ref{eq:master_MB}) gives the condition 
\begin{eqnarray} 
u_{L} (n_{i}-1,n_{i+1}, n_{i+2}+1, n_{i+3}, n_{i+4}) P(\cdots, n_{i}-1, n_{i+1}, n_{i+2}+1, \cdots ,) \cr =
 [~v_{R} (n_{i-2},n_{i-1}, n_{i}, n_{i+1}, n_{i+2}) +
  v_{L} (n_{i-2},n_{i-1}, n_{i}, n_{i+1}, n_{i+2})~] \cr  ~~\times P(\{n_{i}\}). \label{eq:tri_MB_rate}
\end{eqnarray}
\end{enumerate}
One can verify that for this model pair-factorized form of steady state (PFSS) as in Eq. (\ref{eq:tri_Pn}) is indeed possible using MB and Eq. (\ref{eq:tri_MB_rate}) is satisfied when $u_{R}(.) = u_{L}(.) = u(.)$ and  the hop rates  satisfy the  conditions
\begin{eqnarray}
\fl u_{R} (n_{i-2},n_{i-1}, n_{i}, n_{i+1}, n_{i+2}) 
 =u_{L} (n_{i-2},n_{i-1}, n_{i}, n_{i+1}, n_{i+2})]   =
 u (n_{i-2},n_{i-1}, n_{i}, n_{i+1}, n_{i+2})\cr\fl=  \frac{g(n_{i-1}, n_{i}-1)}{g(n_{i-1}, n_{i})}
\frac{g(n_{i}-1, n_{i+1})}{g(n_{i}, n_{i+1})}
 \times \frac{h(n_{i-2}, n_{i}-1)}{h(n_{i-2}, n_{i})}
\frac{h(n_{i}-1, n_{i+2})}{h(n_{i}, n_{i+2})}~~{\rm and}~\\\nonumber\\
\fl [v_{R} (n_{i-2},n_{i-1}, n_{i}, n_{i+1}, n_{i+2}) 
 + v_{L} (n_{i-2},n_{i-1}, n_{i}, n_{i+1}, n_{i+2})]   =
 u (n_{i-2},n_{i-1}, n_{i}, n_{i+1}, n_{i+2}). \cr
 \label{eq:tri_oth_rates}
\end{eqnarray}
\subsection{Calculation of observable in PFSS}

We can express the steady state probability $P(\{n_{i} \})$ following Eq. (\ref{eq:tri_Pn}) as 
\begin{equation}\label{eq:tri_PnF}
 P(\{n_{i}\}) = \frac{1}{Q_{L,N}} \prod_{i=1}^{L} F(n_{i},n_{i+1}, n_{i+2})~ \delta \left(\sum_{i=1}^{N} n_{i} - N\right)
\end{equation}
with $F(n_{i},n_{i+1}, n_{i+2}) = \sqrt{g(n_{i},n_{i+1}) g(n_{i+1},n_{i+2})}~ h(n_{i}, n_{i+2})$ and the canonical partition function as
\begin{equation}
 Q_{L,N} = \prod_{i=1}^{L} F(n_{i},n_{i+1}, n_{i+2})~ \delta \left(\sum_{i=1}^{N} n_{i} - N\right).
\end{equation}
We  can rewrite the expression of hop rate $u (n_{i-2},n_{i-1}, n_{i}, n_{i+1}, n_{i+2})$ in terms of the weight function $F(.)$  following Eqs. (\ref{eq:tri_uR2}), (\ref{eq:tri_PnF})
\begin{eqnarray}
\fl u (n_{i-2},n_{i-1}, n_{i}, n_{i+1}, n_{i+2}) = &\frac{F(n_{i-2}, n_{i-1}, n_{i}-1)}{F(n_{i-2}, n_{i-1}, n_{i})} \frac{F(n_{i-1}, n_{i}-1, n_{i+1})}{F(n_{i-1}, n_{i}, n_{i+1})}& \cr &
 \times \frac{F(n_{i}-1, n_{i+1}, n_{i+2})}{F(n_{i}, n_{i+1}, n_{i+2})}&
\end{eqnarray}
and the hop rates $v_{R} (n_{i-2},n_{i-1}, n_{i}, n_{i+1}, n_{i+2})$ and $v_{L} (n_{i-2},n_{i-1}, n_{i}, n_{i+1}, n_{i+2})$ can be chosen accordingly  that they satisfy Eq. (\ref{eq:tri_oth_rates}). Let us consider that the weight function $F(n_{i},n_{i+1}, n_{i+2})$ can be written by three 2-dimensional representation of matrices \cite{mpa_AC}
\begin{equation}\label{eq:tri_weight_fn}
F(n_{i},n_{i+1}, n_{i+2}) =\langle \alpha (n_{i}) |\Gamma (n_{i+1})| \beta (n_{i+2})\rangle.
\end{equation}
In grand canonical ensemble  following Eq. (\ref{eq:tri_weight_fn}), the partition sum can be written as   $Z_{L}(z) = \sum_{N=0}^{\infty} z^{N} Q_{L,N} = \Tr[T(z)]^{L}$, where $z$ is the fugacity and have a relation with the density of the system as $\rho = z \left(\frac{\partial \ln Z_{L}(z)}{\partial z}\right )$ and the  transfer matrix  be
\begin{equation} \label{eq:tri_transferM}
 T(z) = \sum_{n=0}^{\infty} z^{n} (|\beta (n) \rangle \otimes I) \Gamma (n) (I \otimes \langle\alpha (n)|).
\end{equation}
Here, $I$ be the identity matrix of the same dimension and we used the fact that direct product of any two vectors $|b\rangle$ and $\langle a |$ can be written as 
 $|b\rangle \langle a | = (I \otimes \langle a |) (|b \rangle  \otimes I)$. 
Now, with a simple choice of the hop rates, the weight function $F(n_{i}, n_{i+1}, n_{i+2})$ can be calculated and using transfer matrix  following Eq. (\ref{eq:tri_transferM}), one can in principle calculate the expectation value of any desired observable \cite{FRP, mpa_AC}.

\section{Summary}

The steady states of non-equilibrium systems are very much dependent on the complexity of the dynamics and  it is
difficult to track down a  systematic procedure to obtain the steady state measure
of a system with a given dynamics. In this regard, starting from the Master equation
that governs the time evolution of a many particle system in the configuration space,
several flux cancellation schemes have been in use  for obtaining  the exact  steady state  weight.  These schemes include matrix product ansatz (MPA) \cite{derrida__tasep_mpa}, h-balance scheme \cite{asymm_FRP} and pairwise balance condition (PWB). In this article  we introduced  a new kind of balance condition, namely multibalance (MB), where the sum of  incoming fluxes from  a set of configurations to any configuration $C$  is balanced by the sum of  outgoing fluxes  to  set of configurations chosen suitably.

We have applied  MB condition to a class of nonequilibrium lattice models on a ring where  particles hop to its nearest neighbours and for some cases next to nearest neighbours, with a rate that depends on the
occupation of all the neighbouring sites within a range. We have solved exactly the asymmetric  ZRP   in  two  and  three dimensions and discussed that a factorized steady state (FSS) can be obtained  when hop rates satisfy a  specific condition.  More over, the asymmetric ZRP  in two dimensions  exhibits the phenomena negative differential mobility (NDM) \cite{NDM_PKM}. We have discussed the steady states obtained by MB for asymmetric finite range process (FRP) with nearest neighbours and as well as next nearest neighbours hopping.  It gives us a  steady state in  cluster-factorized  form (CFSS) which helps us in calculating the steady state average of  
the observable using Transfer Matrix method  introduced  earlier \cite{FRP}.

We have  also discussed the two species FRP with directional asymmetry in hop rates and shown that this model  too has a CFSS. The model with $K = 0$ reduces to the
two species zero range process (ZRP) \cite{Evans_Hanney} having a FSS. This two species FRP  having directional asymmetry, with nearest neighbours and next nearest neighbours hopping, can be solved  using the MB condition. The  steady state can be obtained for certain conditions on hop rates and one can calculate the steady state observable here. At the last part  of our article, we have discussed how this balance condition could be applied for other kind of driven interacting many-particle systems. We have another  interesting example,  the 
periodic  triangular lattice models (that we introduced here), where  a particle from a randomly chosen site can hop to one of  its four neighbours with asymmetric rates. MB can be employed  to solve this model exactly  and obtain  a pair factorized steady state (PFSS) under  certain conditions on the hop rates.

We should mention that, we have only tried to formulate a new kind of balance condition to obtain the NESS. We emphasized here
mainly about the application of this balance condition for different kinds of nonequilibrium models and found the conditions of being steady states. One can easily find out the observable such as density,  correlation functions and others at steady state.  More importantly, this method could help in finding the exact steady state structure in models even when the interactions extend beyond two sites.

In summary, we introduced  a new  kind of flux balance condition, namely MB  to obtain steady state   weights  of  nonequilibrium systems  and  demonstrate  its utility in many different kinds  of non-equilibrium dynamics,  including those where the interactions extend beyond two sites.  We   believe that 
the  MB technique will be  very helpful in finding steady state  of many other nonequilibrium systems.

\ack The author would like to gratefully acknowledge P. K. Mohanty for his constant encouragement and careful reading of the manuscript. His insightful and constructive comments have helped a lot in improving this work. The author also acknowledges the support of Council of Scientific and Industrial Research, India in the form of a Research Fellowship 
(Grant No. 09/489(0112)/2019-EMR-I).

\end{document}